\documentclass[conference]{IEEEtran}
\IEEEoverridecommandlockouts
\usepackage{cite}
\usepackage{amsmath,amssymb,amsfonts}
\usepackage{algorithmic}
\usepackage{graphicx}
\usepackage{textcomp}
\usepackage{xcolor}
\usepackage{booktabs}
\usepackage{multirow}
\def\BibTeX{{\rm B\kern-.05em{\sc i\kern-.025em b}\kern-.08em
    T\kern-.1667em\lower.7ex\hbox{E}\kern-.125emX}}
\begin{document}

\title{Decoupled Smart Contract Audits: Lightweight LLM Framework via Distillation and Aggregation}

\author{
\IEEEauthorblockN{
    Bagus Rakadyanto Oktavianto Putra\IEEEauthorrefmark{2},
    Muhamad Risqi Utama Saputra\IEEEauthorrefmark{3},
    Widyawan\IEEEauthorrefmark{2},
    and Guntur Dharma Putra\IEEEauthorrefmark{2}\IEEEauthorrefmark{1}
}

\IEEEauthorblockA{
    \IEEEauthorrefmark{2}Universitas Gadjah Mada, Indonesia and 
    \IEEEauthorrefmark{3}Monash University, Indonesia\\
    bagus.rak2002@mail.ugm.ac.id,
    risqi.saputra@monash.edu,
    widyawan@ugm.ac.id,
    gdputra@ugm.ac.id
}

\thanks{\IEEEauthorrefmark{1}Corresponding author.}
}

\maketitle

\begin{abstract}
Smart contracts face critical security challenges that require thorough auditing in decentralized web services. While Large Language Models (LLMs) have shown promise in automated vulnerability detection, existing approaches lack severity evaluations with actionable remediation and demand unnecessarily massive computational overhead.
In this study, we introduce an efficient end-to-end smart contract security audit framework utilizing lightweight, highly optimized open-source LLMs (0.6B–4B parameters). Our framework decouples comprehensive audit tasks into four interconnected components: vulnerability detection, explanation, severity classification, and remediation recommendation. To maintain high accuracy without massive parameters, we implement Rank-Stabilized Low-Rank Adapters (rsLoRA), knowledge distillation, and a custom Chain-of-Verification (CoVe) aggregation strategy to systematically screen and consolidate multiple draft responses from the model into a highly accurate audit report.
Experimental results demonstrate that our lightweight pipeline consistently outperforms state-of-the-art open-source coder dense LLMs (7B to 34B parameters), achieving 98.25\% accuracy in vulnerability detection and an alignment score of 0.4375 in generative explanation tasks. Furthermore, our extensive ablation studies empirically validate the superiority of our decoupled audit processes over unified prompting and uncover a novel severity centrality bias, establishing a critical benchmark for future research in LLM-assisted auditing.
\end{abstract}


\section{Introduction}
% Blockchain has been gaining traction in various industrial sectors due to the transparency and irreversibility of the blockchain network \cite{khan2021blockchain, habib2022blockchain, lin2022survey}.
In decentralized web services with blockchain as the foundational backbone, smart contracts play an important role in automating the execution and verification of business logic~\cite{lin2022survey}.
% However,  presents unique challenges for smart contract deployments.
However, due to the immutable nature of blockchain, smart contracts cannot be updated with security patches, making them prone to significant vulnerabilities. 
% such as reentrancy attacks and integer overflows.
A notable incident was the DAO hack, which exploited the reentrancy vulnerability and resulted in losses of approximately \$60~million \cite{sayeed2020smart}.
% As validated by the DefiLlama hacks, up to June 2025, the total losses from vulnerability-related attacks reached \$11.7~billion \cite{defillamahacks}. 
Therefore, it is imperative to perform a thorough security audit of the smart contracts prior to their deployment to identify potential security vulnerabilities and prevent financial losses~\cite{he2020smart, david2023you}.

The growing amount and complexity of smart contract projects have led to automated security assessments with Large Language Models (LLMs)~\cite{sun2024gptscan, xia2024auditgpt, zaazaa2024smartllmsentry, wei2024llm_smartaudit, ding2025smartguard, boi2024vulnhunt, yuan2025llmbugscanner}, in which many of them rely on massive, monolithic general-purpose LLMs with unified zero-shot or in-context learning techniques~\cite{sun2024gptscan, ding2025smartguard}. 
% Crucially, these existing monolithic approaches predominantly focus on binary vulnerability detection. They largely overlook the necessity of evaluating vulnerability severity and generating actionable remediation code, leaving developers with alerts that lack prioritization and mitigation strategies.
% In addition, several approaches incorporate integration with traditional program analysis \cite{sun2024gptscan} or static tools \cite{zaazaa2024smartllmsentry}, code alignment with Ethereum Request for Comment (ERC) standards~\cite{xia2024auditgpt}, prompt engineering with in-context learning~\cite{zaazaa2024smartllmsentry, boi2024vulnhunt}, and a combination of Chain-of-Thought (CoT) prompting and in-context learning techniques \cite{ding2025smartguard}. 
% However, these studies~\cite{sun2024gptscan, xia2024auditgpt, zaazaa2024smartllmsentry, wei2024llm_smartaudit, ding2025smartguard, boi2024vulnhunt} have focused exclusively on detecting vulnerabilities in smart contract code without providing explanations or justifications for the identified issues. This lack of interpretability forces developers to manually inspect flagged code, a time‑consuming process that can waste effort on false positives or overlook critical issues due to false negatives.
While these massive models offer strong general reasoning, relying on a monolithic architecture for the whole complex audit workflow is computationally expensive and inefficient. In addition, recent research demonstrates that these large models suffer from \textit{parameter dilution} across highly structured concurrent tasks, whereas small language models with targeted Supervised Fine-Tuning, can significantly outperform models of up to 500 times their size on focused workflows~\cite{jhandi2025slm, belcak2025slm}. Moreover, current security approaches suffer from similar inefficiencies, focusing primarily on binary vulnerability detection, largely overlooking the critical necessity of severity evaluation and actionable remediation~\cite{sun2024gptscan, xia2024auditgpt, zaazaa2024smartllmsentry, wei2024llm_smartaudit, ding2025smartguard, boi2024vulnhunt}. In fact, auditing without severity classification to prioritize developer responses and without specific recommendations to patch the security holes, developers are often overwhelmed by unactionable vulnerability flags~\cite{koszo2025actionable, hu2025suppressed}. 
% Recent research in agentic AI and software engineering has demonstrated that large models suffer from \textit{parameter dilution} when forced to handle varied, highly structured concurrent tasks~\cite{jhandi2025slm, belcak2025slm}.
% Conversely, small language models decoupled and specialized via targeted Supervised Fine-Tuning (SFT) can significantly outperform models up to 500 times their size on focused workflows~\cite{jhandi2025slm, belcak2025slm}. 

Recent studies have begun fine-tuning open-source LLMs specifically to detect or explain vulnerabilities based on curated datasets~\cite{yu2024smart_llama, wei2024ftsmartaudit, ma2024iaudit}. 
% However, these approaches typically utilize a large foundation LLM with 7B to 34B parameters, which cannot be loaded directly on a consumer GPU with limited VRAM.
However, running LLM inference on state-of-the-art coder-dense models (i.e., 7B-34B parameters) requires substantial GPU VRAM, frequently causing out-of-memory errors on standard entry-level to mid-range consumer hardware, which is frequently constrained to 8-12 GB of VRAM~\cite{nvidia_compare_gpus_specs}. Relying on expensive cloud-based GPU instances to run these large baselines creates a significant financial barrier for independent Web3 developers and small open-source teams \cite{belcak2025slm}. To truly democratize smart contract security, auditing tools must be executable on standard consumer hardware. Furthermore, modern software engineering is moving towards \textit{shift-left} security, where vulnerability detection runs locally and continuously within early development and automated testing pipelines~\cite{bhardwaj2024cicd}.
% Massive monolithic models are too resource-intensive for this.
Developing highly optimized, lightweight local models is not just a hardware necessity, but a crucial step towards seamless, localized integration into everyday development workflows.

In this study, we propose an efficient, decoupled end-to-end smart contract security audit framework utilizing a lightweight LLM (i.e., 0.6-4B parameters). In our decoupled smart contract audit, we abandon the traditional monolithic approach, where a single prompt forces one model to balance diverse reasoning tasks simultaneously. Instead, we decouple the workflow into \textit{four distinct specialized modules} that sequentially pass context: vulnerability detection, explanation, severity classification, and remediation. To maintain high detection performance, we adopt targeted Supervised Fine-Tuning utilizing Rank-Stabilized Low-Rank Adapters (rsLoRA)~\cite{kalajdzievski2023rslora} across all sub-tasks, replacing the conventional full fine-tuning. For the complex generative tasks of vulnerability explanation and remediation recommendation, we integrate reasoning-focused Knowledge Distillation~\cite{tian2025llm_distill, hsieh2023llm_distill, wei2022cot, kojima2022zeroshotCoT} from a highly capable reasoning teacher model (Qwen3-30B)~\cite{yang2025qwen3} alongside a custom, modified Chain-of-Verification (CoVe) prompting method~\cite{dhuliawala2024cove}. Here, CoVe mechanism acts as a consensus-driven aggregator to filter, verify, and consolidate multiple independent draft responses generated by the model into a single, highly accurate output.

We divide our smart contract security auditing process into four components, mirroring the way human experts typically conduct audits. First, \textbf{Vulnerability Detector}, is responsible for detecting code vulnerabilities. If vulnerabilities are detected, the second component, namely \textbf{Vulnerability Explanator}, constructs a proof-of-concept implementation with step-by-step explanations on how vulnerabilities can be recreated and exploited, which clearly explains the issues within the code. The third component, \textbf{Severity Determinants}, is designed to assess the severity level of the vulnerabilities, which is critical in prioritizing mitigation strategies. Finally, our last component, the \textbf{Recommender}, provides recommendations on how to fix vulnerability issues. With this interconnected end-to-end components, our framework advances previous approaches which have largely neglected severity evaluation and robust remediation generation.

To objectively assess the validity of our solution, we evaluated our framework using our own dataset by collecting and processing smart contract projects and their corresponding security audit reports. We obtained \textbf{930 vulnerable} code samples and \textbf{969 non-vulnerable} ones, which we refer to as detection dataset. We trained the Vulnerability Detector on this detection dataset to distinguish safe code from vulnerable ones. In addition, we built a separate vulnerability dataset derived from the audit reports from reputable Web3 security platforms (e.g., Code4rena and Shieldify Security). Our vulnerability dataset contains vulnerability descriptions and severity annotations, which categorizes them into \textbf{218 low-severity}, \textbf{549 medium-severity} and \textbf{232 high-severity} vulnerabilities. The vulnerability dataset is used to train our Vulnerability Explanator, Severity Determinant, and Recommender, whose collective goal is to produce end-to-end security audit reports. For comparative evaluation, we benchmarked each task in our framework against a range of state-of-the-art coder-focused large language models, including CodeLlama-13B, CodeLlama-34B, CodeGemma-7B, Codestral-22B, DeepSeek-Coder-7B, DeepSeek-Coder-33B, Qwen2.5-Coder-14B, and Qwen2.5-Coder-32B.

In summary, the main contributions of this paper are described as follows:
\begin{itemize}
    \item We propose an efficient, decoupled end-to-end smart contract security audit framework by dividing the audit process into four specialized sub-tasks, which passes context sequentially: vulnerability detection, explanation, severity classification, and remediation.
    % This establishes that massive parameter counts are not strictly necessary for high-accuracy smart contract auditing.
    \item We introduce an advanced optimization strategy for the complex reasoning tasks of Vulnerability Explanation and Recommendation. Our approach uniquely incorporates reasoning-focused knowledge distillation with a custom, modified Chain-of-Verification (CoVe) prompting method. Our prompting method acts as a consensus-driven aggregator, systematically filtering, verifying, and consolidating multiple candidate analyses into a single, highly accurate audit report.
    \item We pioneer an in-depth investigation into the capacity of open-source LLMs to determine vulnerability severity. We uncover a novel empirical finding: severity classification is significantly more challenging for LLMs than binary vulnerability detection. Furthermore, we demonstrate that current open-source models (both zero-shot and fine-tuned) suffer from a distinct "severity centrality bias," disproportionately defaulting to "Medium" severity while struggling to accurately differentiate "Low" and "High" edge cases. 
    % This finding establishes a critical new benchmark and identifies a major hurdle for future research in automated auditing.
    \item We conducted an extensive evaluation of our proposed framework with ablation studies to systematically quantify the impact of our framework's underlying components. We demonstrated the distinct advantages of our architectural choices and that highly optimized, lightweight open-source LLMs (i.e., 0.6B-4B parameters) consistently outperform larger state-of-the-art open-source coder dense LLMs (i.e., 7B-34B parameters).
    % for example, removing our modified CoVe and knowledge distillation causes generative alignment to collapse from 0.4375 to 0.1111, while standard LoRA severely degrades detection F1-scores compared to rsLoRA.
    % We demonstrated the distinct advantages of Rank-Stabilized LoRA (rsLoRA) over standard LoRA, the critical necessity of voting mechanisms in classification, and the individual versus combined efficacy of knowledge distillation and CoVe aggregation.
    % Crucially, we empirically prove the superiority of our modular, step-by-step pipeline compared to a monolithic "unified" approach, validating that task separation is essential for optimal LLM performance in security auditing.
\end{itemize}

The rest of the paper is organized as follows. Section~\ref{sec:related_work} reviews related work. Section~\ref{sec:data_preprocessing} details our data collection and dataset formation. Section~\ref{sec:llm_system} outlines the proposed decoupled smart contract audit framework. Section~\ref{sec:result} presents our experimental setup, evaluation results, and ablation studies. Section~\ref{sec:threats-to-validity} discusses the threats to validity, and Section~\ref{sec:conclusion} concludes the paper.

% \section{LLMs in Software Engineering: From Code Understanding to Security Auditing}
\section{Related Work}
\label{sec:related_work}
Recent advancements in LLMs have greatly impacted various domains. The introduction of the Transformer architecture~\cite{vaswani2017attention}  marked a paradigm shift, demonstrating that a well-designed attention mechanism alone is sufficient to build a powerful language model. The success of general-purpose Transformers has inspired a transition toward decoder-only architectures optimized specifically for code intelligence. Unlike earlier models, these modern LLMs are trained with code-specific objectives to capture both the syntactic and semantic structures of programming languages. Models like CodeLlama \cite{roziere2023codellama}, DeepSeek-Coder \cite{guo2024deepseek}, Codestral \cite{mistral2024codestral}, CodeGemma \cite{codegemma2024}, and Qwen2.5-Coder \cite{hui2024qwen25coder} represent the current state-of-the-art Dense model in open-source code intelligence, consistently outperforming earlier architectures in standard evaluations by scaling instruction tuning on diverse code data.

The use of LLM in coding tasks extends beyond classification and generation. Several studies have explored their application in auditing code security, particularly for detecting vulnerabilities in software projects \cite{sun2024llm4vuln}. Recent research has further focused on Web3 projects, with Yuqiang et al. \cite{sun2024gptscan} proposing a method called GPTScan for identifying logic vulnerabilities in Solidity-based smart contracts. GPTScan utilizes predefined scenarios and characteristics of vulnerable code, matching them against the target code that needs to be audited. Various studies leverage monolithic LLMs for smart contract auditing such as AuditGPT \cite{xia2024auditgpt}, SmartLLMSentry \cite{zaazaa2024smartllmsentry}, VulnHunt-GPT \cite{boi2024vulnhunt}, LLM-SmartAudit \cite{wei2024llm_smartaudit}, and SmartGuard \cite{ding2025smartguard} utilizing techniques like in-context learning and predefined vulnerable scenarios. Despite demonstrating good performance, all these previous studies rely on massive, monolithic LLMs to handle complex reasoning tasks simultaneously. Recent architectural studies demonstrate that relying on a single large model for coupled reasoning and execution often leads to error propagation, high latency, and format hallucination \cite{deng2025decoupling}. Furthermore, previous research predominantly focuses on binary vulnerability detection, ignoring the critical necessity of explaining the detected vulnerabilities and evaluating their corresponding severity levels.

Given the limitation of general-purpose closed source LLMs, some recent studies have begun utilizing specialized open-source LLMs \cite{yu2024smart_llama, wei2024ftsmartaudit, ma2024iaudit}  not only for vulnerability detection but also for vulnerability explanation. Yu et al. \cite{yu2024smart_llama}  developed Smart-LLaMA which uses two-stage post-training approaches to fine tune an 8B model, enabling their system to both detect and explain the vulnerabilities. Furthermore, Ma et al. \cite{ma2024iaudit}  proposed iAudit based on a 13B model to not only detect smart contract vulnerabilities but also provide justification. The previous research mainly employ a massive LLM with 7-13B parameters, requiring a large amount of computational resources. While high-end enthusiast GPUs have recently expanded to 16-24 GB of VRAM, standard entry-level to mid-range consumer hardware frequently remains constrained to 8-12 GB \cite{nvidia_compare_gpus_specs}. Monolithic models severely push or entirely exceed these limits, severely hindering democratized access and scalable deployment for independent Web3 developers \cite{nvidia_compare_gpus_specs}. Running an LLM with these parameters would certainly increase the chances of an Out-of-Memory (OOM) error, especially when processing a smart contract audit code, which would consume more tokens due to the length of the code.

Recent research has highlighted the efficacy of small language models in resource-constrained environments: when properly fine-tuned, models with small parameters can match or surpass larger models in targeted coding tasks~\cite{hasan2025slm}. Building upon this, we propose a decoupled, lightweight LLM-based smart contract security audit framework. By separating the audit process into specialized modules and employing Knowledge Distillation and a modified Chain-of-Verification (CoVe) aggregation mechanism, our approach completely overcomes the VRAM bottlenecks of the 7B-34B baseline models. Furthermore, our research pioneers the exploration of LLMs for severity classification and establishes a robust pipeline for generating actionable remediation recommendations.

\section{The Detection and Vulnerability Dataset}
\label{sec:data_preprocessing}
% In this section, we detail the construction and utilization of the datasets essential to our smart contract security analysis.
% This section outlines the datasets used to train and evaluate our solution. We begin by detailing our data collection process, which aggregates a comprehensive corpus of vulnerable and benign contracts. We then introduce the Detection Dataset used to train the classification mode and conclude by presenting the Vulnerability Dataset, which facilitates the generation of detailed security reports.

\subsection{Data Collection}
% content: \\
% - Data retrieved from web3bugs \\
% - Preliminary Extraction Steps

\begin{figure}[t]
\centerline{\includegraphics[width=0.48\textwidth]{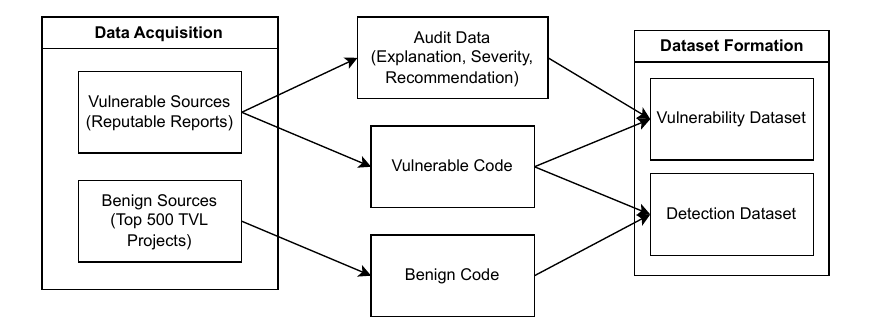}}
\caption{
% \textbf{Data Processing Pipeline.}
The extraction of raw data from audit reports and on-chain sources, the parsing of specific code snippets and metadata, and the subsequent organization into the Detection Dataset (for discrimination tasks) and the Vulnerability Dataset (for generation tasks).}
\label{fig:preprocess}
\end{figure}

% To construct a comprehensive corpus for smart contract security analysis, 
We aggregated data from two primary streams: confirmed vulnerable contracts and benign contracts. As illustrated in Figure \ref{fig:preprocess}, our data processing pipeline transforms these raw sources into structured formats suitable for model training. For vulnerable samples, we crawled publicly available audit reports from reputable smart contract security platforms, specifically Code4rena and Shieldify Security. These sources were selected for their structured reporting standards, which facilitate the precise extraction of "affected code" segments alongside their corresponding vulnerability explanations, severity classifications, and remediation recommendations. For the non-vulnerable/benign samples, we adopted the methodology from prior studies \cite{sun2024gptscan, zaazaa2024smartllmsentry} that takes top N reputable smart contract projects as benign dataset. We use the top 500 smart contract projects by Total Value Locked (TVL) that are currently active on-chain, operating under the assumption that high-value contracts with sustained activity and no reported incidents serve as reliable proxies for secure code samples.

Following the acquisition and parsing of these raw materials, the data is organized into two distinct subsets to support specific experimental objectives:
\begin{enumerate}
    \item \textbf{Detection Dataset}: This dataset contains both the extracted vulnerable code snippets and the benign code samples. The inclusion of benign samples is critical for training the model to discriminate between vulnerable and non-vulnerable code patterns.
    \item \textbf{Vulnerability Dataset}: This dataset is composed exclusively of the vulnerable instances paired with their corresponding metadata (audit reports). It preserves the rich semantic link between the code and its security analysis, enabling tasks related to vulnerability description and report generation.
\end{enumerate}

To rigorously evaluate the Unified vs. Decoupled pipeline architectures (Section \ref{subsec:unified_vs_decoupled}), we constructed an \textbf{aggregated end-to-end dataset}. We applied a strict cross-referencing filtration script to drop any test instances whose base code appeared in the vulnerability training split. This guarantees zero data leakage, ensuring that our models are evaluated on purely unseen, out-of-distribution smart contract logic.
\subsection{Detection Dataset}
The Detection Dataset serves as the foundation for training the \textbf{Vulnerability Detector} to distinguish between vulnerable and non-vulnerable/benign code. A critical challenge in constructing this dataset is the inherent structural disparity between the two classes. Vulnerable samples, extracted from audit reports, typically consist of focused code snippets or isolated functions. In contrast, the raw benign samples are entire smart contract files containing dozens or hundreds of functions (from raw smart contract projects repository).

If left unaddressed, this disparity introduces a confounding variable: the model could learn to associate "short code" with vulnerability and "long code" with safety, rather than learning semantic security features. To mitigate this length bias, we developed a Graph-Based Stratified Sampling algorithm to generate benign samples that statistically mirror the complexity distribution of the vulnerable dataset.

To transform large benign files into comparable code snippets without losing semantic context, we treat each smart contract as a Function Call Graph (FCG). We extract valid connected sub-graphs to generate a massive candidate pool of benign snippets of varying sizes. To mitigate length bias, we applied a graph-based stratified sampling algorithm. By retrieving an equal number of unique benign snippets to match the frequency distribution of total functions in the vulnerable dataset, we minimized the Kullback-Leibler (KL) divergence between the two distributions. This statistical alignment ensures that our model is trained on semantic indicators of vulnerability rather than superficial structural artifacts. Following this balancing process, we partitioned the dataset into training, validation, and testing sets using a 70/15/15 (\%) split. This yielded a final composition of 1,329 training samples, 285 validation samples, and 285 testing samples.

\begin{figure*}[t]
\centering
\includegraphics[width=0.8\textwidth]{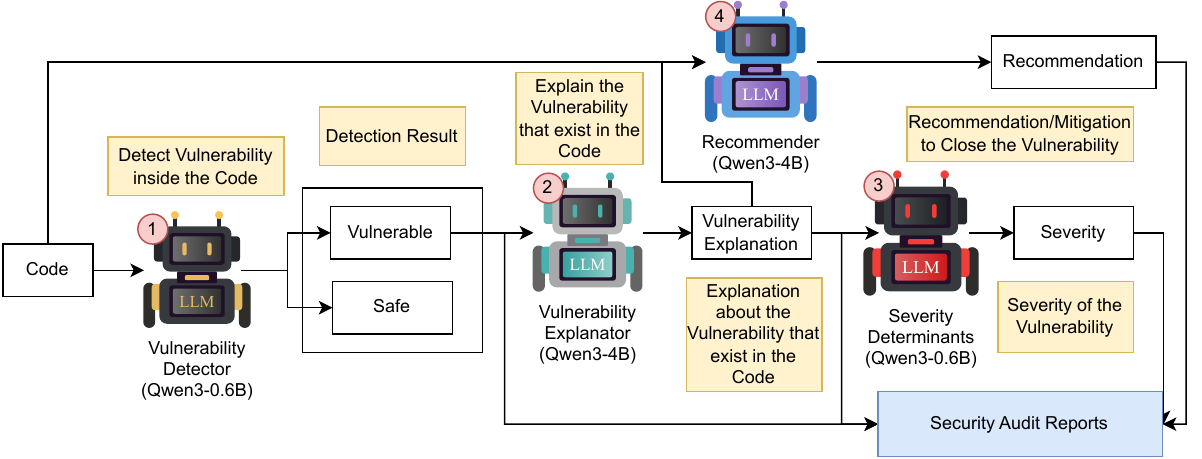}
\caption{The end-to-end workflow of our proposed smart contract security audit framework, which consists of four inter-connected components: 1) Vulnerability Detector, 2) Vulnerability Explanator, 3) Severity Determinants, and 4) Recommender. The outputs from each component will collectively form a complete security audit report.}
% \Description{The overall workflow proposed in this study consists of the detection, explanation, severity determination and recommendation processes.}
\label{fig:llm_system}
\end{figure*}

% \subsection{Data Preparation for Severity Determination}
\subsection{Vulnerability Dataset}
The Vulnerability Dataset is
% serves as the semantic backbone for our automated auditing pipeline, 
designed to facilitate the generation of comprehensive security reports, which consists of confirmed vulnerable code instances paired with their respective audit metadata, including vulnerability explanation, severity, and recommendation steps.
To ensure a rigorous evaluation and prevent data leakage, we implemented a Project-Aware Stratified Splitting strategy. Naively splitting the dataset by individual functions could result in leakage, where code from the same smart contract project appears in both the training and testing sets, artificially inflating performance. To mitigate this, we grouped samples by project identity and employed an iterative stratification algorithm to maintain consistent severity distributions (High, Medium, Low) across partitions while strictly enforcing project-level isolation. This process resulted in a final distribution of 710 training samples, 145 validation samples, and 144 testing samples.

% We leverage this partitioned dataset to train three specialized sub-modules, each targeting a specific component of a professional audit report:
% \begin{enumerate}
%     \item \textbf{Vulnerability Explanator}: We utilize the dataset to train a model that establishes a direct mapping between the raw smart contract code and its natural language vulnerability explanation. By training on these pairs, the model learns to comprehend the normal execution flow and identify specific flaws within that flow, effectively articulating how the code deviates from safe behavior.
%     \item \textbf{Severity Determinant}: The dataset facilitates the training of the Severity Determinant. We design this task to map the vulnerability explanation to a categorical severity label (low, medium, high). This approach mimics the human auditing process, where the assessment of risk magnitude is derived from a contextual understanding of the vulnerability's nature and potential impact.
%     \item \textbf{Recommender}: Finally, we employ the dataset to train the Recommender for synthesizing actionable fixes. This task is formulated as a conditional generation problem where the model is provided with both the original vulnerable code and its explanation. By conditioning the output on both the syntactic structure of the code and the semantic reasoning provided by the explanation, the model is optimized to propose precise code patches that rectify the specific flaw while preserving the contract's intended functionality.
% \end{enumerate}

We leverage this partitioned dataset to train three specialized sub-modules. First, the \textbf{Vulnerability Explanator} is trained to establish a direct mapping between raw smart contract code and its natural language explanation, articulating how the code deviates from safe behavior. Second, the \textbf{Severity Determinant} is trained to map the vulnerability explanation to a categorical severity label (low, medium, high), mimicking the human assessment of risk magnitude based on contextual understanding. Finally, the \textbf{Recommender} is trained as a conditional generation problem; by conditioning the output on both the syntactic structure of the code and the semantic reasoning from the explanation, the model learns to propose precise, actionable code patches that rectify the flaw while preserving intended functionality.

\section{Smart Contract Audit Framework}
\label{sec:llm_system}
% In this section, we introduce our automated Smart Contract Audit Framework, detailing the end-to-end pipeline for vulnerability detection and report generation.

% \subsection{Overview}
Figure \ref{fig:llm_system} depicts our proposed end-to-end smart contract security audit framework utilizing lightweight LLM, particularly Qwen3 with 0.6 and 4B parameters. Based on recent empirical findings \cite{deepscaler2025, jhandi2025slm, hasan2025slm}, open-source LLM with low number of parameters can compete with closed-source LLM with large parameters if the LLM is trained to solve specific tasks. Therefore, instead of using a single monolithic LLM to generate a comprehensive report, our framework decouples the workload across specialized lightweight models. The division of the tasks in our framework also mimic how human experts conduct security audit. As seen in Figure \ref{fig:llm_system}, our framework is divided into 4 components, namely 1) \textbf{Vulnerability Detector}, 2) \textbf{Vulnerability Explanator}, 3) \textbf{Severity Determinants}, and 4) \textbf{Recommender}. Replicating the way human experts typically conduct audits, our framework will first detect vulnerabilities in the smart contract code using the \textbf{Vulnerability Detector} (Section \ref{subsec:detection_n_severity}). Once the vulnerabilities are detected,  the \textbf{Vulnerability Explanator} (Section \ref{subsec:explanator_n_recommender}) will explain and justify the vulnerabilities contained in the code. Afterwards, the \textbf{Severity Determinants} (Section \ref{subsec:detection_n_severity}) will label the severity of the vulnerability contained in the code. Finally, to assist in fixing the security vulnerabilities, the \textbf{Recommender} (Section \ref{subsec:explanator_n_recommender}) will provide recommendations on how to fix the security holes based on the code and the explanations from the Vulnerability Explanator. Our end-to-end framework actually highly depends on the first stage (Vulnerability Detector). It is because only code that is detected as vulnerable by the Vulnerability Detector will proceed to the next stages, which are Vulnerability Explanator, Severity Determinants, and Recommender. Knowing this, evaluating only the first stage will actually show the performance of the end-to-end framework. However, we extend our evaluation by assessing each component using ground truth data.

% To improve the overall model performance and to be comparable or even better with large LLMs, we employ task-aware fine-tuning method and knowledge distillation combined with chain-of-thought (CoT) prompting to train each component of the framework. Task-aware fine-tuning method is utilized to determine the appropriate fine-tuning mechanism used to improve the model performance for each sub-task. Since the vulnerabilities detection and severity determination are considered as classification tasks, an LLM with a very low parameter of 0.6B is used and a full fine-tuning mechanism is employed. For more complex tasks such as vulnerability explanations and recommendations, an LLM model with a parameter of 4B is utilized instead, and a Rank Stabilized
% Low-Rank Adapters (rsLoRA) \cite{kalajdzievski2023rslora} fine-tuning method is employed. The details will be elaborated in the following sections.

To optimize the models across all tasks without the computational burden of full fine-tuning, we strictly utilize Supervised Fine-Tuning (SFT) via Rank-Stabilized Low-Rank Adapters (rsLoRA)~\cite{kalajdzievski2023rslora}. Furthermore, for the complex reasoning tasks (Explanation and Recommendation), we integrate reasoning-focused Knowledge Distillation from a massive teacher model (Qwen3-30B-A3B)~\cite{yang2025qwen3} combined with a custom, modified Chain-of-Verification (CoVe)~\cite{dhuliawala2024cove} aggregation strategy.

% A more detailed explanation of the selection of fine-tuning methods is described in Sections \ref{subsec:detection_n_severity} and \ref{subsec:explanator_n_recommender}. Subsequently, because the Qwen3 model by default supports thinking mode \cite{yang2025qwen3}, in order not to lose its reasoning ability in providing vulnerability explanations and recommendations, the integration technique of knowledge distillation and chain-of-thought prompting is used, which will be discussed in detail in Sections \ref{subsec:knowledge_distillation} and \ref{subsec:explanator_n_recommender}.

% In the audit process, it is generally detected first whether this code has vulnerabilities or not. If it has a vulnerability, then a proof-of-concept or exploitation steps of the existing vulnerability will be made so that an explanation of the vulnerability can be made. Furthermore, the severity of the vulnerability is also determined based on the explanation of the vulnerability in the code to help developers determine the priority of closing security holes. In addition, recommendations for closing security holes are also given to assist developers in patching vulnerabilities.

\subsection{Vulnerability Detector and Severity Determinants}
\label{subsec:detection_n_severity}

% \begin{figure}[t]
% \centering
% \includegraphics[width=\columnwidth]{figure/detector_severity_rev.pdf}
% \caption{The inference phase of our Vulnerability Detector and Severity Determinants, leveraging voting mechanisms via multiple prompts approach.}
% % \Description{Detection and severity process with voting mechanism}
% \label{fig:detector_severity}
% \end{figure}

For both the vulnerability detection and severity determination components, we use Qwen3-0.6B model (0.6 billion parameters) as the backbone LLM. Because these tasks are fundamentally classification problems, a highly optimized 0.6B parameter model is sufficient. We apply rsLoRA training to ensure parameter efficiency while maintaining high accuracy. In addition, in the vulnerability detection and severity determination task, a voting mechanism is also used to enhance the performance of the lightweight LLM such that it can compete with the baseline model. The voting mechanism is implemented using 5 different template prompts provided to the model. In the training phase, each input, either smart contract code (for vulnerability detection) or vulnerability explanations (for severity level determination), is merged with these template prompts and passed through the LLM to learn how to detect vulnerabilities and determine their severity levels. In the inference phase, these template prompts are again utilized together with the inputs and passed through the LLM to produce 5 outcomes which are then used as the voting data. 
% This inference process is described in Figure \ref{fig:detector_severity}.

As for the voting mechanism, it is used to determine the final outcomes of both the vulnerability detection and severity determination. Let \(P = \{p_1, p_2, \dots, p_n\}\) denotes a set of \(n\) distinct prompts utilized during inference. The fine-tuned LLM model processes each prompt \(p_i\) independently, generating a corresponding output \(o_i\) as follows:

\begin{equation}
o_i = \text{LLM}(p_i) \quad \text{for } i = 1, 2, \dots, n.
\label{eqn:voting_output}
\end{equation}

The final output \(O\) is determined through a majority voting process, where we select the most frequently occurring output among the set of predictions \(\{o_1, o_2, \dots, o_n\}\) as computed in the following formula:

\begin{equation}
O = \arg \max_{o} \left(\sum_{i=1}^{n} \mathbb{I}(o_i = o)\right),
\label{eqn:voting_final_output}
\end{equation}

\noindent
where \(\mathbb{I}\) is the indicator function, which returns 1 if the condition is true and 0 otherwise. This voting mechanism leverages the diversity of multiple prompts to enhance prediction reliability. 

To quantify the certainty of the final prediction \(O\), we define a confidence score \(C(O)\) as the normalized count of votes received by \(O\), described as follows:

\begin{equation}
C(O) = \frac{\sum_{i=1}^{n} \mathbb{I}(o_i = O)}{n},
\label{eqn:confidence_score}
\end{equation}

\noindent
where \(\sum_{i=1}^{n} \mathbb{I}(o_i = O)\) counts the number of times \(O\) appears among \(\{o_1, o_2, \dots, o_n\}\). This confidence score reflects the proportion of prompts that agree with the final prediction and serves as an intuitive measure of prediction certainty.

\subsection{Rank Stabilized Low-Rank Adapters}
\label{subsec:rslora}
% content: \\
% - rslora \\ 

To fine-tune the models, instead of leveraging the commonly used Low-Rank Adapters (LoRA) \cite{hu2022lora}, we utilized Rank Stabilized Low-Rank Adapters (rsLoRA) \cite{kalajdzievski2023rslora} since it was demonstrated to be more effective in various tasks. LoRA incorporates an adapter module consisting of two-layer neural networks with low hidden-layer dimensions. However, in standard LoRA, the scaling factor $\gamma_{r}$ is set to $\gamma_{r}=\frac{\alpha}{r}$ for some hyperparameter $\alpha$. This setting is overly aggressive and can cause gradient collapse as the rank increases, slowing the learning process such that using larger ranks yields no better performance than using very small ranks \cite{kalajdzievski2023rslora}. 

To address this, rsLoRA modifies the scaling factor to:
\begin{equation}
    \gamma_{r}=\frac{\alpha}{\sqrt{r}}
\end{equation}

Experimental results, supported by rigorous mathematical proof, show that rsLoRA performs better when trained with larger ranks \cite{kalajdzievski2023rslora}, allowing us to maintain parameter efficiency without sacrificing accuracy.

\subsection{Vulnerability Explanator and Recommender}
\label{subsec:explanator_n_recommender}

% \begin{figure}[t]
% \centering
% \includegraphics[width=\columnwidth]{figure/explanator_recommender_rev-1.pdf}
% \caption{The inference process of Vulnerability Explanator and Recommender which consists of providing candidates answers and aggregating the final answer.}
% \label{fig:explanator_recommender}
% \end{figure}

% \begin{figure}[t]
% \centerline{\includegraphics[width=0.8\columnwidth]{figure/response_prompt.pdf}}
% \caption{Response prompt with CoT data for the Vulnerability Explanator (left) and Recommender (right). The CoT data is generated by Qwen3-14B. This knowledge distillation process was applied in order to maintain the reasoning capabilities of Qwen3-4B that used in Vulnerability Explanator and Recommender}
% \label{fig:response_prompt}
% \end{figure}

We use larger LLM with 4 billion parameters, i.e., Qwen3-4B, to build the Vulnerability Explanator and Recommender, since the task of explaining vulnerabilities and providing recommendations is much more complex than simple detection of vulnerabilities. Furthermore, Qwen3-4B's reasoning capabilities, enhanced by its thinking mode, are also expected to significantly support the generation of vulnerability explanations and recommendations. However, knowledge distillation (described in Section \ref{subsec:knowledge_distillation}) is required to maintain Qwen3-4B's reasoning capabilities in the thinking mode, by training the model to think step-by-step through a predefined CoT prompting.

The Vulnerability Explanator and Recommender consist of two parts: the candidate generator and the final answer aggregator.
% , as described in Figure \ref{fig:explanator_recommender}. 
Note that we only need to train the candidate generator default thinking model is used for the aggregator. The process of generating outputs in both Vulnerability Explanator and Recommender is essentially the same except that each component has different input and output pairs. In the Vulnerability Explanator, the input is a piece of vulnerable code, combined with 5 distinct template prompts. This input is processed by the candidate generator to produce multiple candidate vulnerability explanations. The aggregator then evaluates these candidates to identify the most plausible explanation, which is compiled into a final output. On the other hand, the Recommender's input consists of vulnerable code and the corresponding vulnerability explanation. This input is also combined with 5 different prompts and sent to the Recommender's candidate generator, which produces 5 candidate recommendations for fixing the security flaw. The aggregator then selects the most suitable recommendation as the final output, aimed at resolving the security vulnerability.

\subsection{Chain-of-Verification Aggregation}
To prevent hallucinations and ensure the highest quality output, the candidate responses must be filtered and consolidated. The standard Chain-of-Verification (CoVe) method, introduced by Dhuliawala et al.~\cite{dhuliawala2024cove}, mitigates hallucinations by prompting a model to (1) draft an initial response, (2) plan verification questions to fact-check its draft, (3) answer those questions independently, and (4) generate a final verified response. 

However, standard CoVe suffers from a structural limitation in complex code auditing: it relies on a single initial draft. If the model's initial draft contains a strong, plausible hallucination, it frequently biases the subsequent verification questions, leading to "coupled instability" where the model verifies its own incorrect logic. 

To address this, we introduce a \textbf{Modified CoVe Aggregation} method. Rather than generating a single baseline draft and self-verifying, our Aggregator strictly evaluates the multiple independent candidate responses already produced by the Candidate Generator. Our modified CoVe process is executed through a custom instruction prompt that guides the model through the following internal reasoning steps:

\begin{enumerate}
    \item \textbf{Contextual Grounding:} The model independently reads and traces the provided smart contract code (and the verified vulnerability explanation, in the case of the Recommender) to understand the intended execution flow.
    \item \textbf{Cross-Candidate Verification Planning:} For \textit{each} of the 5 candidate analyses, the model generates specific verification questions to test whether the candidate's claims are mathematically and logically supported by the code.
    \item \textbf{Independent Execution:} The model answers all verification questions to evaluate the factual validity of each candidate, checking if the proposed flaw or mitigation breaks core functionality.
    \item \textbf{Verified Synthesis:} The model identifies the single logical root cause (or mitigation) that fully passes verification. It then consolidates only the verified, accurate elements from the candidates into one final, professional-grade audit report, explicitly discarding speculative or unrelated vulnerabilities.
\end{enumerate}

By decoupling candidate generation from verification and aggregating over multiple independent drafts, our Modified CoVe strictly limits the propagation of hallucinations and ensures high factual alignment with the ground truth.

\subsection{Knowledge Distillation via CoT Prompting}
\label{subsec:knowledge_distillation}

\begin{figure}[t]
\centering
\includegraphics[width=\columnwidth]{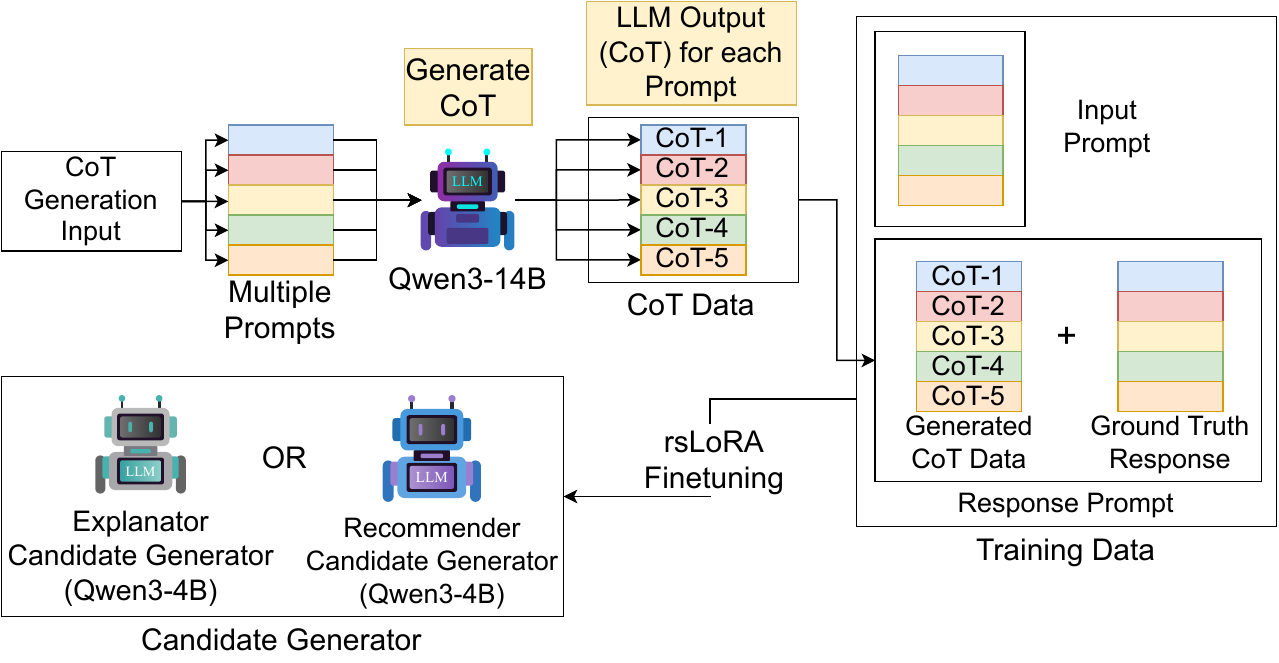}
\caption{Knowledge Distillation process via Chain-of-Thought (CoT) prompting for the Vulnerability Explanator and Recommender.}
% The main idea is the Qwen3-14B will generate the step-by-step that makes a chain-of-thought thinking walkthrough from code to found a vulnerability (for Vulnerability Explanator) and walkthrough from code and vulnerable explanation to giving recommendation (for Recommender)}
% \Description{CoT data generation process with multi prompt and knowledge distillation process}
\label{fig:distillation}
\end{figure}

In developing the Vulnerability Explanator's and Recommender's candidate generator (as discussed in Section \ref{subsec:explanator_n_recommender}), we proposed the use of knowledge distillation to further improve the model performance. Figure \ref{fig:distillation} illustrates the complete process of distilling the reasoning capabilities of a large teacher model (Qwen3-30B-A3B \cite{yang2025qwen3}) to a small student model (Qwen3-4B) by leveraging CoT prompting. This selection was driven by two primary factors. First, Qwen3-30B-A3B demonstrates state-of-the-art logical reasoning capabilities among open-weight models, achieving top-tier performance on the rigorous, contamination-free LiveBench reasoning benchmark~\cite{livebench}. Second, it provides an optimal balance between reasoning prowess and computational feasibility, with its MoE architecture activating only 3B parameters during inference. This ensures our distillation framework remains highly practical and reproducible for researchers with constrained computational resources, without sacrificing the quality of the teacher's reasoning outputs.

For both the Vulnerability Explanator and the Recommender, 5 CoT outputs are generated by the teacher model, which are then combined with the ground truth responses. These are then used, along with the input prompt, to fine-tune the student model using rsLoRA as described in Section \ref{subsec:rslora} and \ref{subsec:explanator_n_recommender}. By incorporating reasoning data in the response prompt through the CoT outputs generated by the larger model, we expect that our smaller model will be able to mimic the reasoning capabilities of the teacher model via strong-to-weak knowledge transfer \cite{yang2025qwen3}.

For Explanator's candidate generator, the teacher model is asked to provide CoT reasoning (e.g., step-by-step explanation) to find security vulnerabilities given the smart contract code and its corresponding vulnerability explanation. As described above, those CoT outcomes will be used as part of the input response to train the Explanator's candidate generator (Qwen3-4B). On the other hand, for Recommender's candidate generator, we also ask the teacher model to provide CoT reasoning steps for fixing security vulnerabilities, based on the smart contract code, its vulnerability explanation, and its recommendations to fix the security problems. Those CoT outputs are then used as part of the input response to train Recommender's candidate generator (Qwen3-4B).

\section{Performance Evaluation}
\label{sec:result}
This section presents a comprehensive evaluation of our framework's performance, assessing both its individual modules and overall systemic efficiency.

\subsection{Baselines and Evaluation Metrics}
\label{subsec:experimental_setup}

While we utilize a highly capable MoE model (Qwen3-30B-A3B) as our teacher for knowledge distillation, our baseline evaluation focuses exclusively on dense architectures. Although MoE models activate fewer parameters during inference, their total memory footprint (VRAM required to load the full weights) remains prohibitive for the resource-constrained consumer hardware targeted by our framework. Therefore, comparing our lightweight models against SOTA dense coder models (7B-34B) provides the most realistic assessment of performance within strict local-deployment memory bounds.

Accordingly, our baseline comprises the most capable open-weights, coder-specialized dense models currently available: CodeLlama (13B, 34B) \cite{roziere2023codellama}, CodeGemma (7B) \cite{codegemma2024}, Codestral (22B) \cite{mistral2024codestral}, DeepSeek-Coder (7B, 33B) \cite{guo2024deepseek}, and Qwen2.5-Coder (14B, 32B) \cite{hui2024qwen25coder}. These models represent the current upper bound of performance for dense coder models in the open-source landscape and serve as the standard against which we measure our proposed lightweight solution. We evaluated these baselines under both Zero-Shot (ZS) and Supervised Fine-Tuned (FT) settings where applicable.

To evaluate the performance of the models and the baselines, we used several metrics depending on the sub-tasks. To evaluate the classification-related task (e.g., vulnerability detection and severity determination) we used accuracy to show the general model performance and \textbf{macro-averaged} precision, recall, and F1-score to treat each label equally. 

To analyze the generative alignment of the Vulnerability Explanator and Recommender with human-authored ground truth audit reports, we shifted away from traditional n-gram metrics, which often struggle to capture deep semantic equivalence and logical correctness in complex technical domains. Instead, building on established prior work \cite{liang2022holistic, zheng2023judging}, we adopted an advanced LLM-as-a-judge approach. This evaluation paradigm has demonstrated significant advantages in assessing textual alignment and has recently been validated specifically for evaluating smart contract vulnerability justifications against ground truth \cite{ma2024iaudit}. Specifically, we utilized the RewardBench-2 benchmark \cite{malik2025rewardbench} leading metric, \textbf{LMUnit 70B}~\cite{saad2024lmunit}, to act as evaluator. Because the native LMUnit prompt lacks an exact template for complex security alignment, we leveraged the \textbf{ATLA-Selene}~\cite{atla2025selene} evaluation prompt schema, specifically adopting their \texttt{classification-with-reference} prompt template. This configuration instructs the LMUnit 70B evaluator to strictly verify the factual and logical consistency between our model's generated output and the ground truth reference report, outputting an alignment score between 0 and 1.
% To analyze the alignment of the Vulnerability Explanator and Recommender with the ground truth or human auditor, we used the LLM evaluator approach inspired by Wei Ma et al. \cite{ma2024iaudit}, who utilized LLMs in analyzing the alignment of vulnerability justification output.  However, our method differs in design and implementation; in particular, we employed Qwen3-14B with “thinking mode” enabled as "the evaluator" since it has good reasoning capabilities and proven to be superior in several benchmarks compared to the current baseline models such as deepseek-r1-qwen-32B and QwQ-32B \cite{yang2025qwen3}. In practice, we prepared a strict prompt that asks the evaluator to compare between the ground truth and the output from our proposed model (or the baselines). The evaluator was instructed to output a value of 1 if both were aligned, and 0 if they were not. To further minimize potential bias and enhance consistency in the evaluation, alongside with strict prompt we also adopted a constrained generation setup (e.g., low decoding temperature and restricted parameterization) to ensure objective responses from the LLM. 
To this end, the LLM alignment value can be calculated as a percentage of evaluator outputs, which can be mathematically represented as follows:

\begin{equation}
\text{Alignment Percentage} = \frac{\sum_{i=1}^{N} x_i}{N} \times 100\%,
\label{eqn:alignment_percentage}
\end{equation}

\noindent
where \(N\) is the total number of evaluator outputs, and \(x_i\) is defined as
\[
x_i =
\begin{cases}
1, & \text{if the output is aligned} \\
0, & \text{if the output is not aligned}.
\end{cases}
\]
Additionally, we also evaluate the computational cost between our proposed method and the baselines. For this experiment, we utilized a \textbf{V100 GPU} with \textbf{32 GB of VRAM}. GPU utilization and maximum used GPU memory were utilized as the computational cost metrics.

\begin{table*}[t]
    \centering
    % Adjust the widths (0.68 and 0.3) if the tables look unbalanced
    \begin{minipage}[b]{0.6\textwidth}
        \caption{Performance Comparison on Vulnerability Detection and Severity Classification}
        \label{tab:classification_results}
        \centering
        \scalebox{0.7}{ % Scaled down slightly more to fit the minipage
        \begin{tabular}{l|rrrr|rrrr}
        \toprule
        \multirow{2}{*}{\textbf{Model}} & \multicolumn{4}{c|}{\textbf{Vulnerability Detection}} & \multicolumn{4}{c}{\textbf{Severity Classification}} \\
        & \textbf{Precision} & \textbf{Recall} & \textbf{F1} & \textbf{Acc} & \textbf{Precision} & \textbf{Recall} & \textbf{F1} & \textbf{Acc} \\
        \midrule
        \multicolumn{9}{c}{\textit{Zero-Shot (ZS) Baselines}} \\
        \midrule
        CodeLlama-13B & 0.6201 & 0.5531 & 0.4753 & 0.5439 & 0.5278 & 0.3639 & 0.1905 & 0.2639 \\
        CodeLlama-34B & 0.7536 & 0.5377 & 0.4066 & 0.5263 & 0.5472 & 0.4776 & 0.3775 & 0.3681 \\
        CodeGemma-7B & 0.6453 & 0.6257 & 0.6095 & 0.6211 & 0.1731 & 0.3221 & 0.1401 & 0.2361 \\
        Codestral-22B & 0.4972 & 0.4982 & 0.4557 & 0.5053 & 0.1852 & 0.3333 & 0.2381 & 0.5556 \\
        DeepSeek-7B & 0.5608 & 0.5464 & 0.5143 & 0.5404 & 0.2365 & 0.3071 & 0.2454 & 0.3333 \\
        DeepSeek-33B & 0.2561 & 0.5000 & 0.3387 & 0.5123 & 0.1852 & 0.3333 & 0.2381 & 0.5556 \\
        Qwen2.5-14B & 0.5255 & 0.5096 & 0.4239 & 0.5193 & 0.2521 & 0.3831 & 0.2558 & 0.3472 \\
        Qwen2.5-32B & 0.7498 & 0.7426 & 0.7389 & 0.7404 & 0.3198 & 0.4596 & 0.3478 & 0.4653 \\
        \midrule
        \multicolumn{9}{c}{\textit{Fine-Tuned (FT) Baselines}} \\
        \midrule
        CodeLlama-13B & 0.9701 & 0.9678 & 0.9683 & 0.9684 & 0.3535 & 0.3672 & 0.3197 & 0.5556 \\
        CodeLlama-34B & 0.9670 & 0.9642 & 0.9648 & 0.9649 & 0.3903 & 0.4064 & 0.3690 & 0.5833 \\
        CodeGemma-7B & 0.9639 & 0.9606 & 0.9613 & 0.9614 & 0.4078 & 0.4738 & 0.4332 & 0.6111 \\
        Codestral-22B & 0.9656 & 0.9645 & 0.9649 & 0.9649 & 0.3880 & 0.4782 & 0.4279 & 0.5903 \\
        DeepSeek-7B & 0.9073 & 0.9044 & 0.9050 & 0.9053 & 0.3678 & 0.3770 & 0.3331 & 0.5625 \\
        DeepSeek-33B & 0.9630 & 0.9608 & 0.9613 & 0.9614 & 0.1852 & 0.3333 & 0.2381 & 0.5556 \\
        Qwen2.5-14B & 0.9688 & 0.9681 & 0.9684 & 0.9684 & 0.3923 & 0.4711 & 0.4260 & 0.5972 \\
        Qwen2.5-32B & 0.9489 & 0.9467 & 0.9472 & 0.9474 & 0.3901 & 0.4613 & 0.4194 & 0.5903 \\
        \midrule
        \multicolumn{9}{c}{\textit{Ours and Ablations}} \\
        \midrule
        \textbf{Ours (rsLoRA + Voting)} & \textbf{0.9829} & \textbf{0.9822} & \textbf{0.9824} & \textbf{0.9825} & \textbf{0.6197} & \textbf{0.5745} & \textbf{0.5877} & \textbf{0.6458} \\
        - w/o voting (rsLoRA) & 0.9755 & 0.9753 & 0.9754 & 0.9754 & 0.3371 & 0.3912 & 0.3521 & 0.5486 \\
        - with voting (LoRA) & 0.9509 & 0.9512 & 0.9509 & 0.9509 & 0.5961 & 0.5280 & 0.5081 & 0.6319 \\
        - w/o voting (LoRA) & 0.9694 & 0.9680 & 0.9684 & 0.9684 & 0.3529 & 0.3559 & 0.2981 & 0.5556 \\
        \bottomrule
        \end{tabular}}
    \end{minipage}
    \hfill
    \begin{minipage}[b]{0.35\textwidth}
        \caption{Alignment Scores for Vulnerability Explanation and Recommendation}
        \label{tab:generative_results}
        \centering
        \scalebox{0.7}{
        \begin{tabular}{lrr}
        \toprule
        \textbf{Model} & \textbf{Vulnerability Explanation} & \textbf{Recommendation} \\
        \midrule
        \multicolumn{3}{c}{\textit{ZS Baselines}} \\
        \midrule
        CodeLlama-13B & 0.055 & 0.250 \\
        CodeLlama-34B & 0.083 & 0.298 \\
        CodeGemma-7B & 0.131 & 0.388 \\
        Codestral-22B & 0.201 & 0.479 \\
        DeepSeek-7B & 0.069 & 0.284 \\
        DeepSeek-33B & 0.229 & 0.333 \\
        Qwen2.5-14B & 0.180 & 0.423 \\
        Qwen2.5-32B & 0.201 & 0.381 \\
        \midrule
        \multicolumn{3}{c}{\textit{FT Baselines}} \\
        \midrule
        CodeLlama-13B & 0.138 & 0.326 \\
        CodeLlama-34B & 0.131 & 0.326 \\
        CodeGemma-7B & 0.020 & 0.166 \\
        Codestral-22B & 0.250 & 0.312 \\
        DeepSeek-7B & 0.048 & 0.291 \\
        DeepSeek-33B & 0.263 & 0.166 \\
        Qwen2.5-14B & 0.284 & 0.375 \\
        Qwen2.5-32B & 0.222 & 0.375 \\
        \midrule
        \multicolumn{3}{c}{\textit{Ours and Ablations}} \\
        \midrule
        \textbf{Ours (Full)} & \textbf{0.437} & \textbf{0.763} \\
        - std LoRA & 0.298 & 0.625 \\
        - CoVe (rsLoRA) & 0.305 & 0.597 \\
        - CoVe (std LoRA) & 0.319 & 0.423 \\
        - Distill (rsLoRA) & 0.256 & 0.270 \\
        - Distill (std LoRA)& 0.319 & 0.333 \\
        - Basic LoRA & 0.090 & 0.263 \\
        - Basic rsLoRA & 0.111 & 0.187 \\
        \bottomrule
        \end{tabular}}
    \end{minipage}
\end{table*}

\subsection{Vulnerability Detector Result}

To evaluate the Vulnerability Detector, we compared our model against both zero-shot and fine-tuned SOTA baselines. As shown in Table \ref{tab:classification_results}, our decoupled rsLoRA framework achieves the highest overall performance, yielding an F1-score of 0.9824 and an accuracy of 98.25\%. Notably, even the best-performing fine-tuned baseline (Qwen2.5-Coder-14B) falls slightly short of our specialized 0.6B parameter model. 

Crucially, our ablation study validates the components of our detection pipeline. Removing the voting mechanism from our rsLoRA model drops the F1-score to 0.9754, demonstrating the efficacy of our multi-prompt ensemble approach. Furthermore, transitioning from rsLoRA to standard LoRA results in a notable performance degradation (F1 drops to 0.9509), proving that rank-stabilization is vital for capturing the intricate patterns of smart contract vulnerabilities. Zero-shot large models performed poorly by comparison, confirming that massive parameter counts cannot substitute for targeted, task-specific adaptation.

\subsection{Vulnerability Explanator Result}

Generating an accurate, professional-grade explanation of a vulnerability requires complex code-to-text reasoning. Table \ref{tab:generative_results} presents the alignment scores computed by the LMUnit 70B judge. Our proposed framework (0.4375) significantly outperforms all fine-tuned and zero-shot dense baseline models, proving that massive parameter scale (e.g., Qwen2.5 32B at 0.2222) is vastly inferior to specialized distillation and aggregation.

Our detailed ablation studies systematically deconstruct the performance gains. Using our modified CoVe Aggregation without distillation yields a score of 0.3055. Using Knowledge Distillation without CoVe Aggregation yields 0.2569. Standard rsLoRA fine-tuning without either enhancement collapses to 0.1111. This proves that while distillation and CoVe independently provide massive improvements, their unique combination in our full pipeline (\textbf{0.4375}) creates a highly synergistic effect, effectively filtering hallucinations generated by the underlying model.

\subsection{Severity Determinants Result}

Determining the severity of a vulnerability requires nuanced contextual reasoning. As shown in Table \ref{tab:classification_results}, our Severity Determinant achieves the highest macro F1-score (0.5877) and accuracy (64.58\%), outperforming all baseline models by a considerable margin. 

Crucially, our empirical investigation into this sub-task reveals a distinct algorithmic phenomenon we term \textbf{Severity Centrality Bias (SCB)}. As observed in the results, baseline models (both zero-shot and fine-tuned) exhibit varying degrees of sensitivity when transitioning from binary vulnerability detection to multi-class severity assessment. Rather than making high-risk extreme predictions on boundary cases, models frequently converge "Low" and "High" severities into the "Medium" class to conservatively manage cross-entropy loss during uncertain predictions.

\begin{figure}[t]
\centerline{\includegraphics[width=0.9\columnwidth]{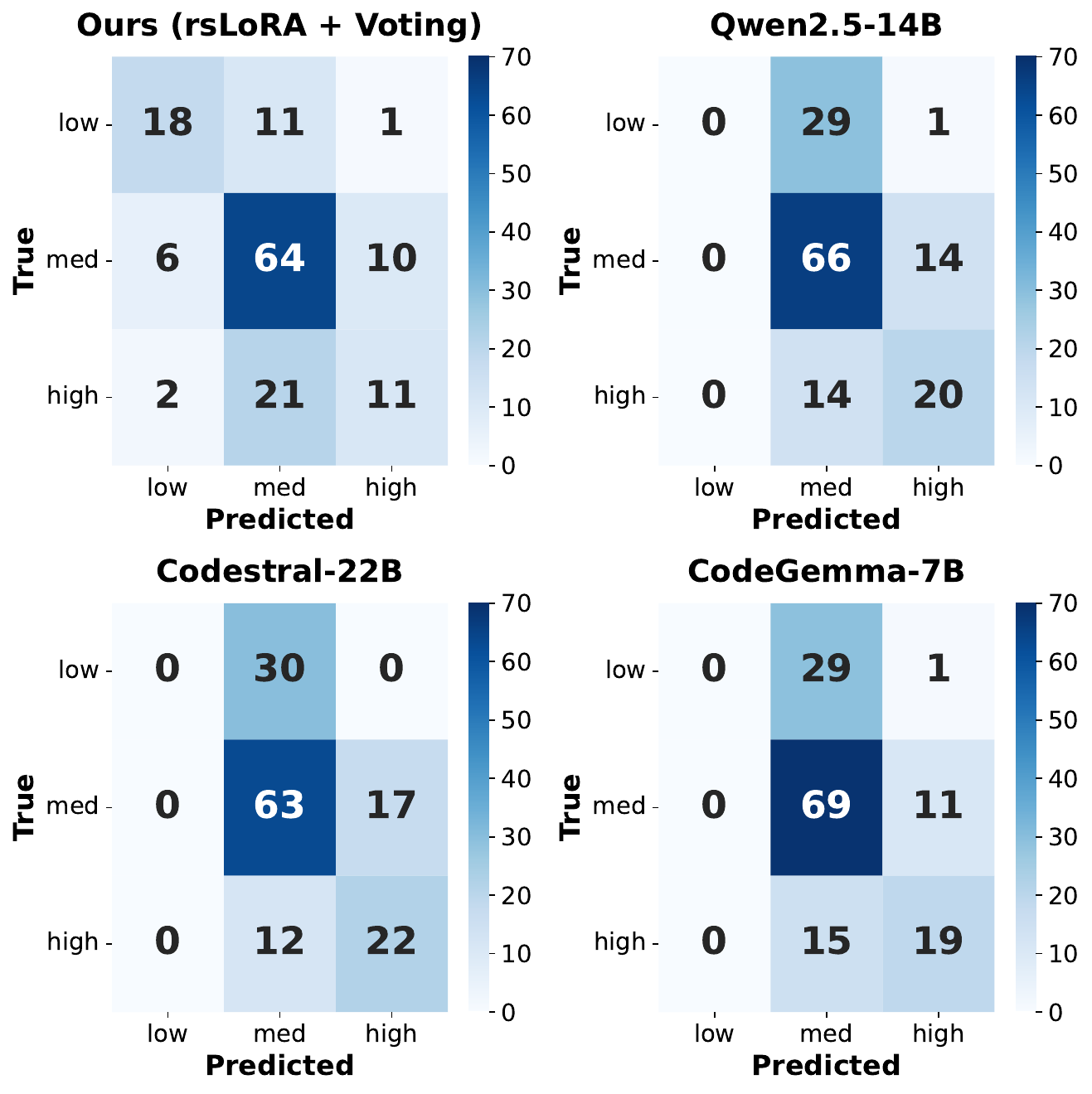}}
\caption{Confusion matrices illustrating the "Severity Centrality Bias" (SCB) across our fine-tuned model and top-performing baselines. While massive baseline models (Qwen2.5-14B, Codestral-22B, CodeGemma-7B) completely fail to identify "Low" severity edge cases and overwhelmingly default to the "Medium" class to minimize extreme predictive penalties, our decoupled framework (top-left) successfully retains multi-class boundary sensitivity.}
\label{fig:severity_conf_matrix}
\end{figure}

To empirically validate this phenomenon across the LLM landscape, we visualize the confusion matrices of our best-performing model alongside three top baseline models (Qwen2.5-14B, Codestral-22B, and CodeGemma-7B) in Figure \ref{fig:severity_conf_matrix}. As the matrices demonstrate, SCB is a pervasive flaw in current code LLMs. Strikingly, the massive baseline models exhibit a complete collapse on the "Low" severity class, misclassifying nearly all true "Low" instances (e.g., 29 or 30 out of 30 cases) directly into the "Medium" category. Similarly, they consistently misassign a large portion of true "High" cases into the "Medium" baseline. In contrast, while our optimized model (rsLoRA + Voting) still exhibits a marginal degree of this centrality bias (conservatively assigning 21 out of 34 true "High" cases and 11 true "Low" cases to "Medium") it successfully preserves boundary sensitivity. Our model accurately identifies 64 out of 80 "Medium" cases, while successfully categorizing 18 "Low" and 11 "High" cases where baseline models entirely fail or severely underperform.

This phenomenon highlights that severity classification is a significantly more challenging reasoning task for LLMs than simple detection. Because "Medium" represents the safest middle ground when a model encounters conflicting contextual features, LLMs overwhelmingly default to this class rather than risking a highly penalized extreme prediction. This finding establishes a critical new benchmark and identifies a major hurdle for future research in automated smart contract auditing.

Furthermore, our ablation results reinforce the superiority of rsLoRA over standard LoRA (F1: 0.5081), and the necessity of our multi-prompt voting mechanism to stabilize predictions in highly ambiguous severity edge-cases.

\subsection{Recommender Result}

Providing actionable mitigation strategies is the final and arguably most critical step of a security audit. As shown in Table \ref{tab:generative_results}, our Recommender achieves an LMUnit 70B alignment score of 0.7639, substantially outperforming all baseline models. The strongest zero-shot baseline, ZS\_Codestral22B, achieves 0.4792, while the best fine-tuned baseline reaches only 0.3750. This highlights the effectiveness of our proposed training framework for generating high-quality remediation recommendations.

Similar to the Explanator, the ablation study highlights the synergistic power of our framework. CoVe Aggregation alone achieves 0.5972, while Knowledge Distillation alone achieves 0.2708. The standard rsLoRA model without our enhancements performs significantly worse (0.1875), indicating that the majority of the performance gains arise from the combination of structured aggregation and distillation.

It is also noteworthy that baseline models generally obtain higher scores on the Recommendation task compared to the Explanation task. This suggests that generating mitigation code snippets is structurally easier for code-oriented LLMs than producing detailed natural-language reasoning about the root causes of vulnerabilities.

% \begin{table}[htbp]
% \caption{Recommender Alignment Scores}
% \label{tab:recom_results}
% \centering
% \large
% \scalebox{0.75}{
% \begin{tabular}{lr}
% \toprule
% \textbf{Method} & \textbf{Alignment Score} \\
% \midrule
% ZS\_Codestral22B & 0.4792 \\
% ZS\_Qwen25\_14B & 0.4236 \\
% ZS\_CodeGemma7B & 0.3889 \\
% ZS\_Qwen25\_32B & 0.3819 \\
% ZS\_DeepSeekCoder33B & 0.3333 \\
% ZS\_CodeLlama34B & 0.2986 \\
% ZS\_DeepSeekCoder7B & 0.2847 \\
% ZS\_CodeLlama13B & 0.2500 \\
% \midrule
% FT\_Qwen25\_14B & 0.3750 \\
% FT\_Qwen25\_32B & 0.3750 \\
% FT\_CodeLlama13B & 0.3264 \\
% FT\_CodeLlama34B & 0.3264 \\
% FT\_Codestral22B & 0.3125 \\
% FT\_DeepSeekCoder7B & 0.2917 \\
% FT\_DeepSeekCoder33B & 0.1667 \\
% FT\_CodeGemma7B & 0.0000 \\
% \midrule
% \textbf{Ours (rsLoRA + CoVe + Distill)} & \textbf{0.7639} \\
% - (Full method, standard LoRA) & 0.6250 \\
% - (CoVe only) & 0.5972 \\
% - (CoVe only, standard LoRA) & 0.4236 \\
% - (Distill only, standard LoRA) & 0.3333 \\
% - (Distill only) & 0.2708 \\
% - (Basic LoRA, no enhancements) & 0.2639 \\
% - (Basic rsLoRA, no enhancements) & 0.1875 \\
% \bottomrule
% \end{tabular}}
% \end{table}

\subsection{Unified vs Decoupled Pipeline Architecture}
\label{subsec:unified_vs_decoupled}

To empirically validate the core hypothesis of our framework, we compared our 4-stage decoupled pipeline against a Unified prompting approach. Crucially, to isolate the impact of the architecture itself, the Unified setup utilizes the exact same Qwen3-4B base model that powers the generative stages (explanation and recommendation) of our decoupled pipeline. However, rather than operating sequentially, this unified baseline is prompted to simultaneously detect the vulnerability, determine its severity, explain the root cause, and output the recommendation in a single inference pass.

As shown in Table \ref{tab:overall_comparison}, forcing a monolithic model to balance diverse, multi-step structural constraints simultaneously leads to severe formatting degradation and logical hallucination. Even when specifically fine-tuned for this joint multi-task objective using the exact same underlying Qwen3-4B foundation, our decoupled approach vastly outperforms the unified strategy across all sub-metrics. This conclusively proves that task separation and modular adaptation are strictly superior for complex software engineering workflows.

\begin{table}[t]
\centering
\caption{Overall Performance Comparison: Ours vs. Unified Approach}
\label{tab:overall_comparison}
\begin{tabular}{llcc}
\toprule
\textbf{Task} & \textbf{Metric} & \textbf{Ours} & \textbf{Unified} \\
\midrule
\multirow{3}{*}{Vulnerability Detection} 
 & Macro Precision & 0.9816 & 0.8433 \\
 & Macro Recall    & 0.9738 & 0.8487 \\
 & Macro F1-Score  & 0.9776 & 0.8459 \\
\midrule
Explanation & Alignment (\%) & 40.91 & 6.82 \\
\midrule
\multirow{3}{*}{Severity Classification} 
 & Macro Precision & 0.2611 & 0.0909 \\
 & Macro Recall    & 0.1396 & 0.3333 \\
 & Macro F1-Score  & 0.1814 & 0.1429 \\
\midrule
Recommendation & Alignment (\%) & 15.91 & 0.00 \\
\bottomrule
\end{tabular}
\end{table}

While Table \ref{tab:overall_comparison} demonstrates the absolute performance gains of our proposed method, evaluating multi-stage LLM outputs strictly on end-to-end absolute metrics obscures the internal dynamics of error propagation. In any sequential generation task, an error in an early stage inherently penalizes downstream metrics. To unpack this, we conduct an Attrition Analysis (Table \ref{tab:error_propagation}) to measure the \textit{conditional yield}, the success rate of a module given that the preceding modules executed correctly.

\begin{table*}[ht]
\centering
\caption{Attrition Analysis: Tracing Error Propagation Across Processing Stages}
\label{tab:error_propagation}
\resizebox{\textwidth}{!}{
\begin{tabular}{l|ccc|ccc}
\toprule
\multirow{2}{*}{\textbf{Processing Stage}} & \multicolumn{3}{c|}{\textbf{Ours}} & \multicolumn{3}{c}{\textbf{Unified (Baseline)}} \\
& \textbf{Count (N)} & \textbf{Absolute Yield} & \textbf{Conditional Yield} & \textbf{Count (N)} & \textbf{Absolute Yield} & \textbf{Conditional Yield} \\
\midrule
0. Ground Truth (Total)   & 44 & 100.0\% & -       & 44 & 100.0\% & - \\
1. Successfully Detected  & 42 & 95.5\%  & 95.5\%  & 34 & 77.3\%  & 77.3\% \\
2. Correctly Explained    & 18 & 40.9\%  & 42.9\%  & 3  & 6.8\%   & 8.8\% \\
3. Correct Severity       & 9  & 20.5\%  & 50.0\%  & 3  & 6.8\%   & 100.0\%$^*$ \\
4. Correct Recommendation & 7  & 15.9\%  & 38.9\%  & 0  & 0.0\%   & 0.0\% \\
\bottomrule
\multicolumn{7}{l}{\footnotesize \textit{$^*$The 100\% conditional yield for the Unified model's severity is an artifact of an extremely low survivor count (N=3).}}
\end{tabular}
}
\end{table*}

The attrition data reveals several critical insights regarding LLM behavior in vulnerability analysis:

\begin{enumerate}
    \item \textbf{Task Interference in Unified Models:} Despite utilizing the same Qwen3-4B foundation and being explicitly fine-tuned for this joint multi-task objective, the unified baseline suffers heavily from cognitive overload. When forced to execute all tasks simultaneously, it experiences a catastrophic drop-off at the explanation stage (yielding only 6.8\% absolute accuracy) and completely collapses by the recommendation stage (0.0\%).
    \item \textbf{Conditional Resilience of the Pipeline:} At first glance, the pipeline's final recommendation alignment of 15.9\% appears modest. However, the conditional yield proves that the recommendation module is highly capable when provided with accurate context. When our pipeline successfully generates a correct explanation, the severity classification succeeds 50.0\% of the time, and the recommendation succeeds 38.9\% of the time. This demonstrates that downstream modules are effectively executing their designated tasks; their absolute scores are merely constrained by the throughput of the preceding stages.
    \item \textbf{Isolating the Bottleneck:} Unlike unified architectures where the origins of errors are obfuscated in a single black-box output, our decoupled approach explicitly identifies the current bottleneck in LLM-based vulnerability analysis: the explanation generation stage (Stage 2). By isolating this structural weakness, our modular framework allows future work to seamlessly swap in stronger explanation models without needing to alter or retrain the highly effective detection and downstream modules.
\end{enumerate}

\subsection{Computational Cost and Resource Efficiency}

Beyond accuracy, a primary motivation for our decoupled framework is deployment feasibility on consumer-grade hardware. Unified, monolithic architectures require loading massive parameter weights into memory simultaneously, creating a strict hardware bottleneck. We evaluated the mean Video RAM (VRAM) consumption of our framework across its four distinct processing stages against baselines. 

As presented in Table \ref{tab:resource_usage}, large unified models (such as Qwen2.5-32B and CodeLlama-34B) demand upwards of 21 to 24 GB of VRAM across all tasks. This completely exceeds the capacity of standard consumer hardware, requiring expensive data-center GPUs (e.g., A100s) or multi-GPU setups. Even smaller 7B to 14B baselines hover near the 12 GB limit.

In contrast, our decoupled approach inherently optimizes resource allocation. For classification tasks (Vulnerability Detection and Severity Classification), our specialized modules consume merely $\sim$3.11 GB of VRAM. Only when generating complex textual reasoning (Explanation and Recommendation) does our framework utilize its heavier generative modules, which still cap at a highly efficient $\sim$8.64 GB. This targeted resource allocation ensures our pipeline comfortably operates entirely within the constraints of modern consumer hardware (under 10 GB VRAM per stage), making State-of-the-Art vulnerability analysis significantly more accessible.

\begin{table}[t]
\centering
\caption{Mean VRAM Usage (GB) Across Processing Stages}
\label{tab:resource_usage}
\begin{tabular}{lrrrr}
\toprule
\textbf{Model} & \textbf{Detect} & \textbf{Explain} & \textbf{Severity} & \textbf{Recommendation} \\
\midrule
CodeGemma     & 11.93 & 10.09 & 11.16 & 11.30 \\
CodeLlama-13B & 19.20 & 14.82 & 16.61 & 16.85 \\
CodeLlama-34B & 22.94 & 21.69 & 22.23 & 21.57 \\
Codestral     & 16.94 & 15.63 & 16.16 & 15.21 \\
DeepSeek-7B   & 11.72 & 11.64 & 10.38 & 11.39 \\
DeepSeek-33B  & 23.58 & 22.51 & 22.62 & 21.89 \\
Qwen2.5-14B   & 13.69 & 12.73 & 13.30 & 12.90 \\
Qwen2.5-32B   & 24.38 & 22.96 & 23.79 & 22.34 \\
\midrule
\textbf{Ours} & \textbf{3.11} & \textbf{8.64} & \textbf{2.77} & \textbf{8.36} \\
\bottomrule
\end{tabular}
\end{table}

\section{Threats to Validity}
\label{sec:threats-to-validity}
Our evaluation process at the explanation and recommendation stages that compares two texts (between ground truth and LLM output) utilizes LLM as an evaluator (LLM as a Judge). The use of LLM for evaluation has also been used in several papers \cite{liang2022holistic,zheng2023judging} and shows its advantages in textual alignment. There has even been previous research that also evaluated the alignment of smart contract vulnerability justifications by LLM against ground truth with LLM evaluator \cite{ma2024iaudit}. LLMs are inherently stochastic and can introduce evaluation bias or inconsistent scoring. To mitigate this risk and enforce strictly deterministic generation, we explicitly constrained the LLM inference parameters, minimizing the decoding temperature and top-p variance. To empirically verify the stability of this configuration, we conducted multiple independent evaluation passes across our entire test set. The resulting alignment percentages exhibited negligible variance across all trials, confirming high inter-trial reliability. While human expert evaluation remains the gold standard for semantic correctness, utilizing a standardized LMUnit metric ensures our alignment scoring is highly reproducible and free from the subjective variance inherent in manual human grading across complex smart contract vulnerabilities.

\section{Conclusion}
\label{sec:conclusion}
We proposed an efficient, decoupled smart contract security audit framework with lightweight LLM. Replicating the human expert workflow, our framework decouples the audit process into four specialized task: Vulnerability Detector, Vulnerability Explanator, Severity Determinant, and Recommender. With rsLoRA and combination of knowledge distillation and a tailored CoVe aggregation strategy, our pipeline consistently outperforms massive monolithic LLMs.
% while maintaining minimal hardware requirements.
We empirically proved that separating complex workflows into specialized tasks is superior to unified monolithic prompting. Furthermore, we pioneered the analysis of LLM-based severity classification and uncovered the \textit{Severity Centrality Bias}, a phenomenon where models conservatively default to "Medium" severity. Future work will focus on exploring another techniques to mitigate SCB and improve boundary-class sensitivity in severity classification.

\section*{Acknowledgment}
\noindent This work was supported by Hibah Penelitian Fundamental (PFR), Ministry of Higher Education, Science, and Technology Indonesia, grant number 048/E5/PG.02.00.PL/2024 - 2679/UN1/DITLIT/PT.01.03/2024.

\bibliographystyle{IEEEtran}
\bibliography{./biblio}

\end{document}